\documentclass[12pt]{article}
  \usepackage{graphicx}
\begin{document}
\input epsf
\def\be{\begin{equation}}
\def\bea{\begin{eqnarray}}
\def\ee{\end{equation}}
\def\eea{\end{eqnarray}}
\def\d{\partial}
\def\la{\lambda}
\def\eps{\epsilon}
\newcommand{\dm}{\begin{displaymath}}
\newcommand{\edm}{\end{displaymath}}
\renewcommand{\b}{\tilde{B}}
\newcommand{\gm}{\Gamma}
\newcommand{\ac}[2]{\ensuremath{\{ #1, #2 \}}}
\renewcommand{\ell}{l}
\def\bb{$\bullet$}

\def\q{\quad}

\def\bn{B_\circ}

\let\a=\alpha \let\b=\beta \let\g=\gamma \let\d=\delta \let\e=\epsilon
\let\z=\zeta \let\c=\chi \let\th=\theta  \let\k=\kappa
\let\l=\lambda \let\m=\mu \let\n=\nu \let\x=\xi \let\r=\rho
\let\s=\sigma \let\t=\tau
\let\vp=\varphi \let\vep=\varepsilon
\let\w=\omega      \let\G=\Gamma \let\D=\Delta \let\Th=\Theta
              \let\P=\Pi 

\renewcommand{\theequation}{\arabic{section}.\arabic{equation}}
\newcommand{\newsection}[1]{\section{#1} \setcounter{equation}{0}}

\def\nn{\nonumber}
\let\bm=\bibitem

\let\pa=\partial

\vspace{20mm}
\begin{center}
{\LARGE  Bound states of KK monopole and momentum}
\\
\vspace{20mm}
{\bf Yogesh K. Srivastava \footnote{yogesh@pacific.mps.ohio-state.edu} \\}

Department of Physics,\\ The Ohio State University,\\ Columbus,
OH 43210, USA\\
\vspace{4mm}
\end{center}
\vspace{10mm}
\thispagestyle{empty}
\begin{abstract}

We construct metrics for multiple Kaluza-Klein monopole-branes carrying travelling waves along one of the isometry directions (not KK monopole fibre) in ten dimensional type IIB supergravity
and relate them via string dualities to two charge Mathur-Lunin metrics.  We find that adding momentum to $N_K$ coincident monopoles 
leads them to separate in the transverse direction into $N_{K}$ single monopoles. Hence the bound state metrics are perfectly smooth, without  $Z_{N_{K}}$ singularities, and correspond to a system with non-zero extension
in the transverse directions. We compare this solution with other solutions with KK monopole.

\end{abstract}
\newpage
\setcounter{page}{1}
\newsection{Introduction}

The Kaluza-Klein(KK)  monopole solution has attracted considerable attention since it was first proposed by Gross and Perry in \cite{grossperry}. 
It is a purely gravitational solution in string theory and one of its obvious attractions is that it is a 
completely regular solution in string theory. Recently, there has been much interest in studying solutions containing 
KK monopole \cite{bena1,ash,bena2}. Also, as recent work shows, it can be used to connect black rings in five dimensions
to black holes in four dimensions \cite{gaiot,elvang1}. Studies of black rings in Taub-NUT space \cite{bena1,bena2}
led to supersymmetric solutions carrying angular momentum in four dimensional asymptotically flat space \cite{elvang1}. 
KK monopoles also occur in $4$-dimensional string theoretic black holes.

In the past few years, there has been substantial progress in 
constructing microstate solutions for black holes in $5$-dimensions. For the case of $2$-charge systems, all bosonic solutions have been constructed by Mathur and Lunin \cite{lm4,Lunin} . 
Geometries with both bosonic and fermionic condensates were considered in \cite{taylor1} and relationship between gravity and CFT sides has been further 
explored in \cite{taylor2,taylor3,alday1,alday2} recently. Our understanding of the $3$-charge systems is less complete but a few examples are known \cite{Lunin2,gms1,gms2,bena2,levi}. In four dimensions, smooth solutions for $3$ and $4$-charges have appeared in 
the literature \cite{bena1,ash,Bala} .

In this work, we want to study gravity solutions corresponding to KK monopoles carrying momentum. This 
would correspond to a simple system of $2$-charges in $4$-dimensions and would be the first example of such a metric. Note that this solution cannot be obtained by setting one of the charges in the known three charge solultion to zero and U-dualizing. For example, setting D$5$ charge to zero in D$1$-D$5$-KK solution of \cite{bena2}, we get D$1$-KK which can be U-dualized to KK-P.
When we try to put one charge to zero in the geometry of \cite{bena2}, one finds that it reduces to the `naive' 2-charge geometry and on dualization, it gives the
naive KK-P geometry. Here, by `naive' we mean geometries obtained by applying the harmonic-superposition rule. The black-ring structure of the geometry is destroyed when one of the charges (other than the KK monopole) is set to zero. This 
raises the question whether this geometry has all three charges bound and whether this 3-charge system is `symmetric' between the charges. One of 
the motivation for the present work is to understand, in a simplified setting, if the solution constructed in \cite{bena2} is a true bound 
state or not. Our construction of KK-P is manifestly bound and if it can be related by dualities to the solution of Bena and Kraus (with one charge set to zero)
then, at least in this simplified setting, we can be confident that this is a bound state.

Note that in this system we add momentum along one of the 
isometry directions, different from KK monopole fibre direction. Hence this system is still supersymmetric and is not dual to $D0-D6$ system as studied
in \cite{horowitz} which was non-supersymmetric and would correspond to momentum along fibre direction.

Since all two charge systems are related by string dualities, one may ask the reason for constructing KK-P \emph{ab initio} when it can obtained by dualities from F1-P. We will also construct it by dualities from F1-P solution constructed
in \cite{lm4} in section $2$. The reason we also obtain it using Garfinkle-Vachaspati transformation is that it gives us unsmeared solutions which carry $t$ and $y$ 
dependence, $y$ being the direction of wave. We show complete smoothness of this $N$-monopole solution carrying momentum. When we try to get a solution 
independent of $t,y$ by smearing then we will see that singularities develop which are similar to  singularities in solution obtained via dualities. Since
number of KK monopoles is always discrete (even classically), we know that singularities are an artifact of smearing and discrete solution is always
smooth (even classically). One particular feature of these solutions is that orbifold singularities of multiple KK monopoles are also resolved and they are
completely smooth.

\subsection{Outline of the paper}

The plan for present paper is as follows.
\begin{itemize}
\item In $\S1$, we add momentum to KK monopole by the method of Garfinkle-Vachaspati (GV)  transform. 

\item In  $\S2$, we concentrate on the smoothness of $N$ monopoles solution. Specifically, we consider the case of two monopole
solution with momentum. We demonstrate how KK monopoles get separated by the addition of momentum and discuss the 
regularity of solution.

\item In  $\S3$, we get the same solution as above by performing dualities on general two charge solutions constructed in \cite{lm4}.

\item In $\S4$, we  perform T-duality to convert this to KK-F1 solution. 

\item In $\S5$, we consider the KK-D1-D5 metric obtained by Bena and Kraus in the near-horizon
limit and try to see if it is duality symmetric. It turns out that it is not. This is not surprising as Buscher duality rules used are valid only at the supergravity level and as 
 mentioned earlier and discussed in \cite{bena2}, more refined duality rules will be required. 

\item We give our T-duality conventions and a discussion of Garfinkle-Vachaspati (GV) transform in two appendices.
\end{itemize}

\newsection{Adding momentum to KK monopoles by GV transformation}

In this section, we take the metric of a single KK monopole and add momentum to it along one of isometry directions (not the fibre direction) using 
the procedure of Garfinkle and Vachaspati.  Using the linearity of various harmonic functions appearing in metric, we can superpose harmonic functions
to get multi-monopole metric with momentum. 

\subsection{ KK monopole metric}

Ten dimensional metric for KK monopole at origin is 
\be
ds^{2} = -dt^{2} +dy^{2} + \sum_{i=6}^{9}dz^{i}dz_{i} + H[ ds + \chi_{j}dx^{j}]^{2} + H^{-1}[dr^{2} + r^{2}(d\theta^{2} +
\sin^{2}\theta d\phi^{2})]
\ee
\be
H^{-1}= 1+ \frac{Q_{K}}{r} \ \ , \ \ \vec{\nabla} \times \vec{\chi} = - \vec{\nabla}H^{-1}
\ee
Here $y$ is compact with radius $R_{5}$ while
$x_{j}$ with $j=1,2,3$ are transverse coordinates while $z_{i}$ with $i=6,7,8,9$ are coordinates for torus $T^{4}$. Here 
$Q_K = \frac{1}{2} N_K R_K$ where $N_K$ corresponds to number of KK monopoles. Near $r=0$, $s$ circle shrinks to zero. For $N_K =1$, it does so smoothly
while $N_K >1$, there are $Z_{N_K}$ singularities. First we consider just $N_K =1$ case. 
Introducing the null coordinates $u= t+y$ and $v=t-y$, above metric reads
\be
ds^{2} = -dudv + \sum_{i=6}^{9}dz^{i}dz_{i} + H[ ds + \chi_{j}dx^{j}]^{2} + H^{-1}[dr^{2} + r^{2}(d\theta^{2} 
+\sin^{2}\theta d\phi^{2})]
\ee
We want to add momentum to this using Garfinkle-Vachaspati (GV) transform method \cite{garfinkle1,garfinkle2}. 

\subsection{ Applying the GV transform} 

Given a space-time with metric $g_{\mu\nu}$ satisfying the Einstein equations and a null, killing and hypersurface orthogonal vector field $k_{\mu}$ i.e. satisfying the following properties 
\begin{equation}
 \  k^{\mu}k_{\mu} = 0,\  k_{\mu ; \nu}+ k_{\nu;\mu} =0,\ k_{\mu;\nu}= \frac{1}{2}(k_{\mu}A_{,\nu}-k_{\nu}A_{,\mu}) 
\end{equation}
for some scalar function $A$ is some scalar function, one can construct a new exact solution of the equations of motion by defining  
\be
g'_{\mu\nu}= g_{\mu\nu} + e^{A}\Phi k_{\mu}k_{\nu}
\ee
The new metric $g'_{\mu\nu}$ describes a gravitational wave on the background of the original metric provided the matter fields if any. satisfy some conditions \cite{myers} and
the function $\Phi$ satisfies
\be
\nabla^{2}\Phi =0 \ , \ k^{\mu}\partial_{\mu}\Phi =0 
\ee
Some more details about Garfinkle-Vachaspati transform are given in appendix. Note that all this is 
in Einstein frame but it can be rephrased in string frame very easily. In our case, there are no matter fields and the dilaton is zero so there is no difference between the string and Einstein frames. We take 
$(\frac{\partial}{\partial u})^{\mu}$ as our null, killing vector. Since $g_{uu}=0$, it is obviously null and since the metric
coefficients do not depend on $u$, it is also killing. One can also check that this vector field is hypersurface orthogonal for a constant $A$ which may be absorbed in $\Phi$.   
Applying the transform we get
\be
ds^{2} = -(dudv +T(v,\vec{x})dv^{2}) + \sum_{i=6}^{9}dz^{i}dz_{i} + H[ ds + \chi_{j}dx^{j}]^{2} + H^{-1}
[\sum_{j=1}^{3}dx_{j}^{2})]
\ee
where $T(v,\vec{x})$ satisfies the three dimensional Laplace equation. General solution for $T$ is
\be
T(v,\vec{x})= \sum_{\ell \geq 0}\sum_{m=-\ell}^{\ell} [a_{\ell}(v)r^{\ell} + b_{\ell}(v)r^{-\ell +1}]Y_{\ell m}
\ee
Here $Y_{\ell m}$ are the usual spherical harmonics in three dimensions. Constant terms can be removed by a change of coordinates. To see this, we consider $T(v,\vec{x})= g(v)Y_{0m}$. We can 
go to a new set of coordinates $du'= du - g(v)Y_{0m}dv$ and other coordinates remaining same. If we want \footnote{We are excluding vibrations along
the fibre direction. One could include such excitations but making the corresponding solution asymptotically flat turns out to be difficult. Perhaps a formalism different than GV
transform might be better suited for that purpose} a regular (at origin)
and asymptotically flat solution (after dimensional reduction along the fibre) then the only 
surviving term is $T(v,\vec{x})= \vec{f}(v)\cdot \vec{x}$. This is apparently not asymptotically flat but can be made so by the following coordinate transformations
\begin{eqnarray}
v&=& v' \\
\vec{x}&=&\vec{x'} -\vec{F}  \\
u &=& u' -2\dot{F_{i}}x'_{i} + 2\dot{F_{i}}F_{i} - \int^{v'} \dot{F}^{2}(v)dv 
\end{eqnarray}

Here $\vec{f}(v)= -2\ddot{\vec{F}}$ and dot refers to derivative with respect to $v$. Making this  change of coordinates, the terms in metric change as follows
\begin{eqnarray}
dudv = du'dv' -2\dot{F_{i}}dx'_{i}dv' + \dot{F}^{2}(v')dv'^{2} \ \ \\
dx_{j}dx_{j} = dx'_{j}dx'_{j} + \dot{F}^{2}(v')dv'^{2} -2\dot{F_{i}}dx'_{i} dv' \ \
\end{eqnarray}

So the final metric is 
\begin{displaymath}
ds^{2} = -du'dv' + 2\dot{F}_{i}(1-H^{-1})dx'_{i}dv'  -\dot{F}^{2}(1-H^{-1})dv'^{2} + \sum_{i=6}^{9}dz^{i}dz_{i} 
\end{displaymath}
\be
 +   H [ ds + \chi'_{j}dx'_{j} -\chi'_{j}\dot{F}_{j}dv ]^{2} +H^{-1}[\sum_{j=1}^{3}dx_{j}^{'2})]
\ee
Removing the primes,we write the above metric in the form of chiral-null model as 
\be
ds^{2} = -dudv + dz_{i}dz_{i} + H^{-1}dx_{j}^{2} + H(ds + V_{j}dx^{j} + Bdv)^{2} +2A_{j}dx_{j}dv + Kdv^{2}
\ee 
Here we have introduced the notation
\begin{eqnarray}
H^{-1} = 1+ \frac{Q_{K}}{|\vec{x}-\vec{F}(v)|^{2}} \ \ \ \ \ \ \ \ \ \ \ \ \ \ \ \ \ \ \ \  B= -\vec{\chi}\cdot \vec{F}(v)\\
K(x,v)= \frac{Q_{K} |\dot{\vec{F}(v)}|^{2}}{|\vec{x}-\vec{F}(v)|^{2}} \ \ , \ \ A_{i}= 
-\frac{Q_{K} \dot{F}_{i}(v)}{|\vec{x}-\vec{F}(v)|^{2}} \\
\chi_{1} = -\frac{Q_{K}(x_{2}-F_{2}(v))}{(x_{1}-F_{1}(v))^{2}+(x_{2}-F_{2}(v))^{2}}\left(\frac{(x_{3}
-F_{3}(v))}{|\vec{x}-\vec{F}(v)|}\right)  \\
\chi_{2} = \frac{Q_{K}(x_{1}-F_{1}(v))}{(x_{1}-F_{1}(v))^{2}+(x_{2}-F_{2}(v))^{2}}
\left(\frac{(x_{3}-F_{3}(v))}{|\vec{x}-\vec{F}(v)|}\right)
\end{eqnarray}

We have written harmonic functions above for the case of single KK monopole.
But because of linearity, we can superpose the harmonic functions to get the metric for the 
multi-monopole solution, with each monopole carrying it's wave profile $\vec{F}^{(p)}(v)$ and having a charge
$Q^{(p)}= \frac{Q_{K}}{N_{K}}$. Functions appearing in 
the metric then become
\begin{eqnarray}
H^{-1} = 1+ \sum_{p}\frac{Q^{(p)}_{K}}{|\vec{x}-\vec{F}^{(p)}|} \ \ \ \ \ \ \ \ \ \ \ \ \ \ \ \ \ \ \ \ \\
K(x,v)= \sum_{p}\frac{Q^{(p)}_{K} |\dot{\vec{F}^{(p)}}|^{2}}{|\vec{x}-\vec{F}^{(p)}|} \ \ , \ \ A_{i}= 
-\sum_{p}\frac{Q^{(p)}_{K} \dot{F}^{(p)}_{i}}{|\vec{x}-\vec{F}^{(p)}|} 
\end{eqnarray}

Besides the above functions, there are  $\vec{\chi}$ and $\chi_{i} \dot{F}_{i} $ with 
\begin{eqnarray}
\chi_{1} = -\sum_{p}\frac{Q^{(p)}_{K}(x_{2}-F^{(p)}_{2})}{(x_{1}-F^{(p)}_{1})^{2}+(x_{2}-F^{(p)}_{2})^{2}}\left(\frac{(x_{3}
-F^{(p)}_{3})}{|\vec{x}-\vec{F}^{(p)}|}\right)  \\
\chi_{2} = \sum_{p}\frac{Q_{K}^{(p)}(x_{1}-F^{(p)}_{1})}{(x_{1}-F^{(p)}_{1})^{2}+(x_{2}-F^{(p)}_{2})^{2}}\left(\frac{(x_{3}-F^{(p)}_{3})}
{|\vec{x}-\vec{F}^{(p)}|}\right)
\end{eqnarray}  

\newsection {Smoothness of solutions} 
To show smoothness, we concentrate on simple case of two monopoles. So in this section, we consider the simple case of two monopoles carrying waves. Normally (i.e without momentum), one would
expect the system of two monopoles two have orbifold type $Z_{2}$ singularities. But since momentum is expected to separate
the monopoles, this solution would be smooth, without any singularities. As we saw earlier, the metric
for a single monopole carrying a wave is 
\begin{eqnarray}
ds^{2} &=& -dudv + 2\dot{F}_{i}(1-H^{-1})dx_{i}dv  -\dot{F}^{2}(1-H^{-1})dv^{2} + \sum_{i=1}^{4}dz^{i}dz_{i}  \nonumber \\
       & & +   H [ ds + \chi_{j}dx_{j} -\chi_{j}\dot{F}_{j}dv ]^{2} +H^{-1}[\sum_{j=1}^{3}dx_{j}^{'2})]
\end{eqnarray}
After the change of coordinates, we have
\be
H^{-1}= 1+ \frac{Q_{K}}{|\vec{x}-\vec{F}(v)|} 
\ee
For a single monopole, $Q_{K}=\frac{R_{K}}{2}$. 
For two monopoles, we take profile function with $F(v)$ with range from $0$ to $4\pi R_{5}$ where $R_{5}$ is the radius
of $y$ circle. From $0$ to $2\pi R_{5}$ it gives profile function $F_{1}(v)$ for the first monopole while from 
$2\pi R_{5}$ to $4\pi R_{5}$ it gives profile function $F_{2}(v)$ for second monopole. For two monopoles, harmonic 
functions need to be superposed. So we have
\be
H^{-1}= 1+ \frac{Q_{K}}{|\vec{x}-\vec{F}^{(1) }(v)|} + \frac{Q_{K}}{|\vec{x}-\vec{F}^{(2) }(v)|} 
\ee
Since $\vec{\nabla} \times \vec{\chi} = -\vec{\nabla} H^{-1}$ is a linear equation, the function $\vec{\chi}$ also gets superposed and 
\be
\vec{\chi}= \vec{\chi}^{(1)} + \vec{\chi}^{(2)}
\ee
Profile functions $F_{1}$ and $F_{2}$ are given in terms of a single profile function in the covering space
$F(v)$ which goes from $0$ to $4\pi R$ such that
\begin{eqnarray}
F^{(1)}(v)= F(v) \ \mbox{for $v=[0,2\pi R]$} \\
F^{(2)}(v)= F(v -2\pi R) \ \mbox{for $v=[2\pi R,4\pi R]$} \\
F^{(1)}(v=2\pi R)= F^{(2)}(v=0)
\end{eqnarray}

\begin{figure}[t] 
   \begin{center}
   \includegraphics[width=3in]{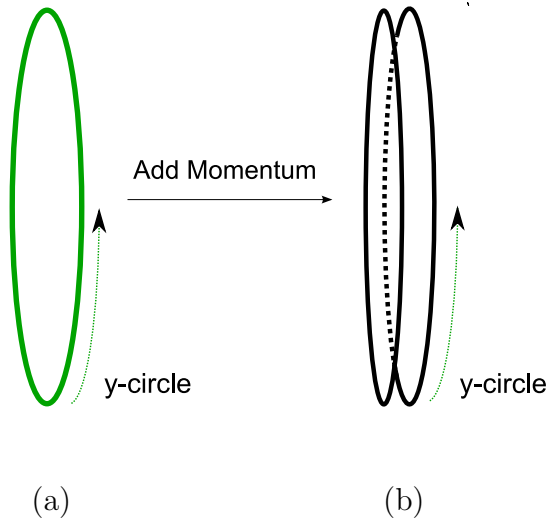} 
    \vspace{.2truecm}
    \end{center}
    \vspace{.2truecm}
   \hspace{4.35truecm}(a)\hspace{4.15truecm}(b)\hspace{3.8truecm}
     \caption{Monopole-strings i.e KK monopoles reduced on $T^4$ (a) 2 coincident monopole-strings(b) 2 single monopole-string separated in transverse directions.} 
  \label{fig:monopole}
\end{figure}

Notice that since one monopole goes right after the other $Q_{K}$ is same for both parts of the harmonic function 
and equal to $Q_{K}$ for single KK monopole. To check the regularity of the two monopole solution, we make the following
observations. Apparent singularities are at the locations $x=F^{(1)}(v)$ and $x=F^{(2)}(v)$. We can go near any one of
them and it is like a single KK monopole (containing terms which do not contain $dv$ or $dv^{2}$) and hence 
smooth. Notice that it is important that poles are not at the same location to avoid conical defects. Locally, we can make 
a coordinate transformation
\begin{eqnarray}
u'= u+ f(x_{i},v)  \ \ \mbox{so that} \ du'= du + \partial_{i}fdx^{i}+ \partial_{v}f dv \\
-dudv + 2A_{i}dx^{i}dv + Kdv^{2}= dv(-du + 2A_{i}dx^{i}+ Kdv)= -du'dv
\end{eqnarray}

by suitably choosing $f(x_{i},v)$. Since such a coordinate transformation can always by locally done, we will only see
single KK monopole which is smooth. Basically, momentum separates a monopole with $N$-unit of charge into $N$
 monopoles of unit charge, each of which is  smooth. If we go near any one, we see only that monopole. For the same reason of monopole 
separation due to momentum, $N$-monopoles with momentum are also  smooth.

\newsection{ Continuous distribution of monopoles}

In this section, we get $t,y$ independent solution by smearing over $v$ which corresponds to metric for multiple KK monopoles distributed continuously. 
But since we have a three dimensional base space, smearing over $v$ gives elliptic function. To see this we 
use three dimensional spherical polar coordinates
\be
x_{1}=\tilde{r}\sin\tilde{\theta}\cos\phi \ , \ x_{2}=\tilde{r}\sin\tilde{\theta}\sin\phi \ , \ x_{3}= \tilde{r}\cos\tilde{
\theta} 
\ee
and following profile function
\be
F_{1} = F\cos(\omega v +\alpha) \ , \ F_{2} = F\sin(\omega v + \alpha)
\ee
This is the profile function used for simplest metric for D$1$-D$5$ system. It is possible that choosing different profile function may 
lead to regular behavior. But the point is that for D$1$-D$5$ system, all metrics (for generic profile functions) were regular while that will not be 
the case here. The smeared harmonic function would be 
\be
H^{-1}= 1+ \frac{Q_{K}}{2\pi}\int_{0}^{2\pi}\frac{d\alpha}{|\vec{x}-\vec{F}|}=1+\int_{0}^{2\pi} \frac{d\alpha}{\sqrt{\tilde{r}^{2} + 
F^{2} -2F\tilde{r}\sin\tilde{\theta}\cos(\omega v + \alpha -\phi)}}
\ee
Using the periodicity of the integral, this reduces to 
\be
H^{-1}= 1+ \frac{Q_{K}}{2\pi}\int_{0}^{2\pi} \frac{d\beta}{\sqrt{\tilde{r}^{2} + F^{2} -2F\tilde{r}\sin \tilde{\theta}\cos\beta}}
\ee
To do the integral, we switch from $\tilde{r},\tilde{\theta}$ to coordinates $r,\theta$ which are defined by
\be
\tilde{r}^{2}= r^{2}+ F^{2}\sin^{2}\theta \ \ , \ \ \tilde{r}\cos\tilde{\theta}= r\cos\theta \ \ , \ \ \tilde{r}^{2}
\sin^{2}\tilde{\theta}= (r^{2}+ F^{2})\sin^{2}\theta
\ee
Using these, we write
\be
H^{-1}= 1+ \frac{Q_{K}}{2\pi}\frac{1}{\sqrt{r^{2}+F^{2}}}\int_{0}^{2\pi} \frac{d\beta}{\sqrt{1+ \frac{F^{2}sin^{2}\theta}
{r^{2}+F^{2}}  -\frac{2F\sin\theta\cos\beta}{\sqrt{r^{2}+F^{2}}}}}
\ee
Writing $p= \frac{F\sin\theta}{\sqrt{r^{2}+F^{2}}}$, we get
\be
H^{-1}= 1+ \frac{Q_{K}}{\pi}\frac{K(p)}{\sqrt{r^{2}+F^{2}}}
\ee
where $K(p)$ is elliptic integral of the first kind and
\be
2K(p)= \int_{0}^{2\pi} \frac{d\beta}{\sqrt{1+p^{2}-2p\cos\beta}}
\ee
$K(p)$ diverges when $p=1$ i.e 
\be
r^{2}+F^{2}= F^{2}\sin^{2}\theta \ \ \mbox{or} \ \ r^{2}+ F^{2}\cos^{2}\theta =0
\ee  
which is the same place where there is an apparent singularity in the geometry of \cite{bal,mm}.  It is known that elliptic integral $K(p)$ diverges logarithmically 
as $p \rightarrow 1$. One can, of course, add a suitable harmonic counterterm to cancel the singularity but then the solution will not be 
asymptotically flat.  Integral for function $K$ appearing in the metric is very similar and it gives 
\be
K= \frac{Q_{K}F^{2}\omega^{2}K(p)}{\pi\sqrt{r^{2}+F^{2}}}
\ee
Similarly, components of $A_{i}$ are given by
\be
A_{\phi}= \frac{2Q_{K}}{\pi}\sqrt{r^{2}+F^{2}}\left(\frac{K(p)-E(p)}{F}\right)
\ee
with other components zero. This is in untilde coordinates. Expressions for functions $\chi_1$, $\chi_2 $ can be obtained by solving the equation 
$\vec{\nabla} \times \vec{\chi} = - \vec{\nabla}H^{-1}$. In the next section, we would connect the above functions to functions obtained 
by dualizing D$1$-D$5$ system. 

\subsection{Singularities}
Due to the presence of elliptic functions and their attendant singularities
the solution above, in the smeared case, is not smooth. In section $2$, we saw that solution with two KK-monopoles with momentum added is 
 smooth. The calculation goes through for $N$-monopole case. This is similar to case of fundamental string and momentum system \cite{lm4} where adding
momentum leads to separation of previously coincident strings. One can ask, what causes singularities to develop in the case when a continuum of
KK monopoles carry momentum. The reason is that harmonic functions like $H^{-1}$ corresponding to three-dimensional transverse space go like $1/r$ and one further
integration (for smearing) effectively converts them into harmonic functions in a two-dimensional transverse space \footnote{One may guess that if we had allowed
vibrations along fibre direction, harmonic functions would be different and smoothness would be maintained even after smearing}  which are known to diverge logarithmically. Elliptic integral $K(p)$, for example,
 also diverges logarithmically as $p$ goes to $1$.

Physically also, we can see that smeared case which corresponds to a continuous distribution of
KK monopoles is expected to have troubles. Normally, we can consider continuous distribution of sources like branes, fundamental string etc in 
constructing metrics in supergravity approximation. Discreteness emerges when we use quantization conditions from our knowledge of string theory sources
and BPS condition. KK monopole solution is different because here discreteness is inbuilt as smoothness of single KK monopole forces definite periodicity 
for compact direction and gives $Q_K = \frac{1}{2} N_k R_K $. So considering a continuous distribution of KK monopoles can give singularities even in cases where
where situation is smooth for large but discrete distribution of KK monopoles. Even though continuous solution is not smooth, it is 
still less singular than `naive' KK-P solution. `Naive' solution 
  is
  \be
  ds^2 = -dt^2 + dy^2 + \frac{k}{r} (dt +dy)^2 + ds_{T^4} + H[ ds + \chi_{j}dx^{j}]^{2} + H^{-1}[dr^{2} + r^{2}(d\theta^{2} +
\sin^{2}\theta d\phi^{2})] 
  \ee
  In smeared solution, we have logarithmic singularity due to elliptic integrals occurring in solution. We note that those singularities are milder than 
what we get in `naive' solution. 
\newsection{ Connecting F-P to KK-P via dualities}

In this section, we connect KK-P metric found above to 2-charge metrics constructed in \cite{lm5} by doing various 
dualities. This will also help in interpreting various quantities appearing in the metric. We start with F1-P metric, 
written in the form of chiral null model. Introducing the null coordinates $u= t+y$ and $v=t-y$, the metric reads

\begin{equation}
ds^{2} = H\left(-dudv +Kdv^{2} + 2A_{i}dx_{i}dv\right) + dx_{i}dx_{i} + dz_{j}dz_{j}
\end{equation}
\begin{eqnarray}
& & B_{uv}= -\frac{(H-1)}{2}, \ B_{vi}= HA_{i},\  e^{-2\Phi}= H^{-1} = 1+ \frac{Q}{|\vec{x}-\vec{F}|^{2}}  \\
& & K(x,v)= \frac{Q |\dot{\vec{F}}|^{2}}{|\vec{x}-\vec{F}|^{2}} \ \ , \ \ A_{i}= -\frac{Q \dot{F}_{i}}{|\vec{x}-\vec{F}|^{2}}
\end{eqnarray}
Here $y$ is compact with radius $R_{5}$ while
$x_{i}$ with $j=1,2,3,4$ are transverse coordinates while $z_{j}$ with $i=6,7,8,9$ are coordinates for torus $T^{4}$.
Summation over repeated indices is implied. We have written harmonic functions above for the case of single string.
But because of the linearity of chiral null model, we can superpose the harmonic functions to the metric for the 
multi-wound string, with each strand carrying its wave profile. Functions appearing in the metric then become
\begin{eqnarray}
H^{-1} = 1+ \sum_{p}\frac{Q^{(p)}}{|\vec{x}-\vec{F}^{(p)}|^{2}} \ \ \ \ \ \ \ \ \ \ \ \ \ \ \ \ \ \ \ \ \\
K(x,v)= \sum_{p}\frac{Q^{(p)} |\dot{\vec{F}^{(p)}}|^{2}}{|\vec{x}-\vec{F}^{(p)}|^{2}} \ \ , \ \ A_{i}= 
-\sum_{p}\frac{Q^{(p)} \dot{F}^{(p)}_{i}}{|\vec{x}-\vec{F}^{(p)}|^{2}}
\end{eqnarray}

Here we have smeared along torus directions so that nothing depends on these coordinates and these are isometry 
directions along which T-duality can be performed. To go from fundamental string carrying momentum (FP) system to 
KK-P system we perform following chain of dualities
\begin{displaymath}
F(y)P(y) \stackrel{S}{\rightarrow} D1(y)P(y) \stackrel{T6789}{\rightarrow} D5(y6789)P(y) \stackrel{S}{\rightarrow}
 NS5(y6789)P(y) \stackrel{T4}{\rightarrow} KK(4y6789)P(y)
\end{displaymath}

In the above we start with type IIB theory and metric above is in string frame. In the final step we will need to smear 
along $x_{4}$ direction so that harmonic functions becomes $3$ dimensional harmonic functions for KK monopole. 
Direction $x_{4}=s$ which is now compact becomes non-trivially fibred with other non-compact directions to give 
KK monopole metric. 
To perform S-duality we first need to go to Einstein frame and there the effect of 
S-duality is to reverse the sign of dilaton and B-field going to RR field. For NS fields the net effect in string 
frame is that dilaton changes sign and metric gets multiplied by $e^{-\Phi}= H^{-1/2}$. Also B-field becomes RR field.
Now we apply four T-dualities along $z_{i}$ directions for $i=6,7,8,9$. 
Since there is no B-field here, only change is in the metric along torus directions. 
In the absence of B-field, RR field only picks up extra indices. So we have following fields for D5-P system.

\begin{eqnarray}
ds^{2} = H^{1/2}\left(-dudv +Kdv^{2} + 2A_{i}dx_{i}dv\right) + H^{-1/2}dx_{i}dx_{i} + dz_{j}dz_{j} \\
e^{-2\Phi}= H^{-1} \ \ , \ \ \ C_{uv6789} = -\frac{(H-1)}{2} \ , \ \ \ \  C_{vi6789}= HA_{i} \ \ \ \
\end{eqnarray}

We need to dualize this $6$ form field to $2$ form field using this metric. First we write down field strengths 
corresponding to above RR fields.
\begin{eqnarray}
G_{uv6789i} &=& \partial_{u}C_{v6789i}+ (-1)^{6}\partial_{v}C_{6789iu}+  ....  + (-1)^{6}
\partial_{i}C_{uv6789} =  -\frac{1}{2}\partial_{i}H \\
G_{vi6789j} &=& \partial_{v}C_{i6789j}+ (-1)^{6}\partial_{i}C_{6789jv}+   ....  + (-1)^{6}
\partial_{j}C_{vi6789} = \partial_{j}(HA_{i}) -\partial_{i}(HA_{j}) \nonumber \\
\end{eqnarray}

Here we have used the fact that direction $u$ and torus directions are isometries. To dualize this we use
\be
G^{\mu_{1}...\mu_{p+1}}= \frac{\epsilon^{\mu_{1}...\mu_{p+1}\nu_{1}...\nu_{9-p}}}{(9-p)!\sqrt{-g}}G_{\nu_{1}...\nu_{9-p}}
\ee
and we normalize $\epsilon$ by $\epsilon^{tyjkli6789}=1$. For our metric we have $\sqrt{-g}= \sqrt{H}$. Also, in terms of 
lightcone coordinates, our epsilon tensor is normalized as $\epsilon^{uvjkli6789}=-2$. Using these, we get dual 3-form
 field strengths.
\begin{eqnarray}
G^{jkl}= \frac{\epsilon^{jkluv6789i}}{7!\sqrt{H}}G_{uv6789i}=\frac{\epsilon^{jkli}\partial_{i}H}{\sqrt{H}} \\
G^{ukl}= \frac{\epsilon^{uklvi6789j}}{7!\sqrt{H}}G_{vi6789j}=\frac{-2\epsilon^{klij}[\partial_{j}(HA_{i}) -
\partial_{i}(HA_{j})]} {\sqrt{H}} 
\end{eqnarray}

we have also reduced ten dimensional epsilon symbol to epsilon tensor in four flat Euclidean dimensions. Now we can use 
metric to lower the indices. We get
\be
G_{mnp}= g_{mj}g_{nk}g_{lp}G^{jkl}= \frac{\epsilon_{mnpi}\partial^{i}H}{H^{2}}= -\epsilon_{mnpi}\partial^{i}H^{-1}
\ee
\begin{equation}
G_{vmn}= g_{uv}g_{mk}g_{nl}G^{ukl}+ g_{vj}g_{mk}g_{nl}G^{jkl}= \epsilon_{mnij}\partial^{i}A^{j}
\end{equation}

where we have flat Euclidean metric in four dimensional space. After this we perform an S-duality to get to NS5-P system.
\begin{eqnarray}
ds^{2} &=& \left(-dudv +Kdv^{2} + 2A_{i}dx_{i}dv\right) + H^{-1}dx_{i}dx_{i} + dz_{j}dz_{j} \\
e^{2\Phi}&=& H^{-1} 
\end{eqnarray}

Under $S$-duality, RR field go to NS-NS B-field. To go to KK monopole we apply a T-duality along a direction 
perpendicular to NS5. Let us choose that to be $x^{4}=s$. Rightnow, $x^{4}$ is not an isometry direction since $H$,
$A_{i}$ and $K$ depend on $x^{4}$. To remedy this we smear along $x^{4}$ direction. Smearing converts four dimensional 
harmonic functions into three dimensional harmonic functions. Now we do T-duality using the Buscher T-duality rules
given in the appendix. We can write the T-dual metric in general as
\be
ds^{2}_{T}= ds'^{2} -\frac{G_{\mu s}G_{\nu s}dx^{\mu}dx^{\nu}}{G_{ss}} + \frac{(ds + B_{\mu s})^{2}}{G_{ss}}
\ee
where $ds'^{2}$ is the original metric minus the $G_{ss}$ part. 
 Then metric in generic form for $KK-P$ (since we do not completely know the values of B-field yet) is
\be
ds^{2} = -dudv + dz_{i}dz_{i} + H^{-1}dx_{i}^{2} + H(ds + B_{\mu s}dx^{\mu})^{2} +2A_{j}dx_{j}dv + (K - HA_{s}^{2})dv^{2}
\ee 
After duality we also have $B'_{\mu s} = HG_{\mu s}$. Since KK-P  is a purely gravitational solution we do not want any 
$B$-field. So we must dualize in a direction in which $\dot{F}_{j}$ is zero. So $G_{\mu s}=0$ and $A_{s}=0$. Since we have
smeared along $x_{4}=s$ direction, any derivatives along $s$ give zero. For field strengths of $NS5-P$, we had
\be
G_{ijk} = -\epsilon_{ijkl}\partial^{l}H^{-1} \ , \ G_{vij} = \epsilon_{ijkl}\partial^{k}A^{l} 
\ee
Consider $l=s$ case first. Since we have smeared along $s$, we get $G_{ijk}=0$ and hence $B^{(2)}=B_{ij}$ becomes pure
gauge after smearing and can be set to zero. So to have non-zero field strength, we must have one $s$ index. Suppose
$i=s$. Then using isometry along $s$, we have
\be
G_{sjk}= \partial_{s}B_{jk}+ \partial_{k}B_{sj}+ \partial_{j}B_{ks}= \partial_{k}\chi_{j}-\partial_{j}\chi_{k}
\ee
where $\chi_{j}=B_{sj}$ is a three dimensional vector field. We see that
\be
\vec{\nabla} \times \vec{\chi} = -\vec{\nabla} H^{-1}
\ee
where $\vec{\nabla} $ is three-dimensional gradient. From other components of field strength, we get
\be
G_{vij}= \partial_{v}B_{ij}+ \partial_{j}B_{vi} + \partial_{i}B_{jv}= \epsilon_{ijkl}\partial^{k}A^{l} 
\ee
Since $A^{s}=0$ and derivative with respect to $s$ gives zero, we have, for 3-dimensional indices $i,j$  
\be
G_{vij}= \partial_{v}B_{ij}+ \partial_{j}B_{vi} + \partial_{i}B_{jv} = (dC)_{ij} =0
\ee
where first term is zero because as we saw earlier, $B_{ij}$ with both indices on 3-dim. space are zero (upto gauge 
transformations). Here $B_{iv}=C_{i}$ is a three dimensional vector field which is again gauge equivalent to zero as can be
seen above. So one of the indices $i,j$ must be $s$ to get no-zero right hand side. Then we get
\be
G_{vis}= \partial_{v}B_{is}+ \partial_{s}B_{vi} + \partial_{i}B_{sv}= \partial_{i}B - \partial_{v}\chi_{i}=
\epsilon_{iskl}\partial^{k}A^{l}= -\epsilon_{ikl}\partial^{k}A^{l} 
\ee
where $B_{sv}= B$ is a three dimensional scalar which satisfies above equation. 
So KK-P metric is 
\be
ds^{2} = -dudv + dz_{i}dz_{i} + H^{-1}dx_{j}^{2} + H(ds + \chi_{j}dx^{j} + Bdv)^{2} +2A_{j}dx_{j}dv + Kdv^{2}
\ee 
Here $i=1,2,3$ and all harmonic functions are in three dimensions. $z_{i}$ for $i=6, ..9$ are torus coordinates. 
Using Garfinkle-Vachaspati transform, we got $B= -\vec{\chi}\cdot \dot{\vec{F}}$. Let us check that it satisfies the 
equation for $B$ written above. By using the rules of three dimensional vector calculus we have
\be
\nabla(\vec{\chi}\cdot \dot{\vec{F}})= (\dot{\vec{F}}\cdot \nabla)\vec{\chi}+(\vec{\chi}\cdot \nabla)\dot{\vec{F}} 
+ \dot{\vec{F}} \times (\nabla \times \vec{\chi}) + \vec{\chi} \times (\nabla \times \dot{\vec{F}})=(\dot{\vec{F}}\cdot 
\nabla)\vec{\chi}+ \dot{\vec{F}} \times (\nabla \times \vec{\chi}) 
\ee
where we have used that $\vec{F}$ only depends on $v$. Comparing this with equation for $B$, we see that it is same 
 when we realize that 
\begin{displaymath}
\vec{\nabla} \times \vec{\chi}  = -\vec{\nabla}  H^{-1} \ \ ,  \ \ A_{j}= (1-H^{-1})\dot{F}_{j}
\end{displaymath}

Only non-trivial step is to show that
\begin{displaymath}
(\dot{\vec{F}}\cdot \nabla)\vec{\chi}= -\partial_{v}\vec{\chi}
\end{displaymath}

To see this we write
\begin{displaymath}
\partial_{v}= \dot{F}_{i}\partial_{F_{i}}= - \dot{F}_{i}\nabla_{i}
\end{displaymath}

where we have used the fact that derivatives with respect to observation point $x_{i}$ and source $F_{i}$ can be 
interchanged at the cost of a minus sign. One can reduce the expression for $B$ to a quadrature but it is not easy to carry out the integration
explicitly. For completeness, we give the formal expression below. 
\be
B= \frac{Q_K}{2} r\cos\theta \int_{0}^{1} g dg \int_{0} ^{2\pi} d\alpha \frac{1}{(r^2 + F^2 g^2 - 2Fgr\sin\theta \cos\alpha)^{3/2}}
\ee
\newsection{Properties of solution}

In this section, we discuss some properties of the new solutions. 

\begin{enumerate}

\item{Smoothness}: As we mentioned earlier, N-monopole solutions are  smooth. $Z_{n}$ singularities associated with coincident 
KK monopoles are lifted by adding momentum carrying gravitational wave. System behaves as $N_{K}$ single monopoles and is  smooth. 
But as we saw earlier, smeared solution has singularities. From the analysis of two monopole case, the reason is apparent. When we consider continuum of KK monopoles, any two monopoles come arbitrarily close
to each other and separation due to momentum is not enough to prevent singularities. Even though logarithmic singularity encountered is quite mild, it 
is not removable by coordinate identification, as was possible in single KK monopole case.

\item{KK electric charge}: In the metric
\be
ds^{2} = -dudv + dz_{i}dz_{i} + H^{-1}dx_{j}^{2} + H(ds + \chi_{j}dx^{j} + Bdv)^{2} +2A_{j}dx_{j}dv + Kdv^{2}
\ee
we have term $B$ which corresponds to momentum along fibre direction $s$ even though we started with no profile-function component along the fibre. 
On dimensional reduction, this gives a KK electric field, in addition, to magnetic field due to KK monopole. Adding momentum to KK monopoles has caused 
this electric field. To see its origin from other duality related systems, note that $B= B_{sv}$ where $B_{sv}$ is the component of B-field in the 
fibre direction. Physically, the angular momentum present in $5$-dimensional metric (NS$5$-P) becomes momentum along fibre direction and manifests itself
as electric field in reduced theory. 
\end{enumerate}

\newsection{Comparison to recent works}

\subsection{Work of Bena-Kraus}

Recently, Bena and Kraus \cite{bena2} constructed a metric for D1-D5-KK system which, according to them, corresponds
to a microstate. These solutions are smooth and are related to similar studies of black rings with Taub-NUT space
 as the base space \cite{bena2,elvang1}. In these metrics, KK monopole charge is separated from D1 and D5
charges and is treated differently from the other two charges.
In this section, we want to check whether Bena-Kraus metric for D1-D5-KK is symmetric under permutation of charges by
duality. For simplicity, we consider near horizon limit of Bena-Kraus(BK) metric and perform dualities to permute
 the charges. BK metric and gauge field (after correcting the typos), in the near horizon limit, is
\begin{eqnarray}
ds^{2}= \frac{1}{\sqrt{Z_{1}Z_{5}}}\left[-(dt+k)^{2} + (dy-k-s)^{2}\right]+ \sqrt{Z_{1}Z_{5}}ds^{2}_{KK} + \sqrt{\frac{
Z_{1}}{Z_{5}}}ds^{2}_{T^{4}} \\
k= \frac{l^{2}}{4\Sigma}\frac{\Sigma -r-\tilde{R}}{Q_{K}}\left(d\psi -\frac{Q_{K}}{R_{K}}d\phi \right) \  \ 
s= -\frac{l^{2}}{2\Sigma Q_{K}}\left((\Sigma -r)d\psi + \frac{Q_{K}}{R_{K}}\tilde{R}d\phi \right) \\
ds^{2}_{KK}= Z_{K}(dr^{2}+ r^{2}d\theta^{2}+r^{2}\sin^{2}\theta d\phi^{2})+ \frac{1}{Z_{K}}(R_{K}d\psi + Q_{K}\cos\theta d
\phi)^{2}
\end{eqnarray}

Here we have used the notation
\begin{eqnarray}
\Sigma = \sqrt{r^{2}+ \tilde{R}^{2}+ 2\tilde{R}r\cos\theta} \ \ , \ \ \tilde{R}= \frac{R_{K}^{2}}{4Q_{K}} \\
Z_{K}= \frac{Q_{K}}{r} \ \ , \ \  Z_{1,5}= \frac{Q_{1,5}}{\Sigma} \ \ , \ \ l^{2}= 4Q_{K}\sqrt{Q_{1}Q_{5}}
\end{eqnarray}

RR two form field is given by
\begin{displaymath}
C^{(2)}= \frac{1}{Q_{1}}\left((r-\tilde{R})dt \wedge dy -\frac{l^{2}}{4}(\tilde{R}+ \alpha) \left[dt \wedge (\frac{d\psi}{
Q_{K}} + \frac{d\phi}{R_{K}}) -dy \wedge (\frac{d\psi}{Q_{K}} - \frac{d\phi}{R_{K}})\right] \right.
\end{displaymath}
\be
 \left. - \frac{(\tilde{R}+ \alpha)
l^{4}}{4Q_{K}R_{K}} d\psi \wedge d\phi \right)
\ee 
Here we have used the notation $\alpha= \Sigma -r$. Bena-Kraus notation $Z_{K}$ is our $H^{-1}$. Their $R_{K}\psi$ is
our $s$ coordinate. Correspondingly, periodicity of $\psi$ is $2\pi$ while period of $s$ was $2\pi R_{K}$.
Dilaton is given by $e^{2\Phi}= \frac{Q_{1}}{Q_{5}}$. Fields given above are for $D1_{y}D5_{y6789}KK_{\psi y 6789}$
 system. 

\subsection{Dualities}
We can perform an S-duality to go to F1-NS5-KK system and then perform a T-duality along the fibre direction
$\psi$ permute the charges of KK monopole and NS5 brane. Since dualities map near horizon region of one metric
to near horizon region of other metric, we expect an interchange of KK and 5-brane charges. 
 After S-duality, we get the following metric for F1-NS5-KK system.
\be
ds^{2}= \frac{1}{\sqrt{Z_{1}Z_{5}}}\left[-(dt+k)^{2} + (dy-k-s)^{2}\right]+ \sqrt{Z_{1}Z_{5}}ds^{2}_{KK} + \sqrt{\frac{
Z_{1}}{Z_{5}}}ds^{2}_{T^{4}} 
\ee
NS-NS two form field is given by
\begin{displaymath}
B^{(2)}= \frac{1}{Q_{1}}\left((r-\tilde{R})dt \wedge dy -\frac{l^{2}}{4}(\tilde{R}+ \alpha) \left[dt \wedge (\frac{d\psi}{
Q_{K}} + \frac{d\phi}{R_{K}}) -dy \wedge (\frac{d\psi}{Q_{K}} - \frac{d\phi}{R_{K}})\right] \right.
\end{displaymath}
\be
 \left. - \frac{(\tilde{R}+ \alpha)
l^{4}}{4Q_{K}R_{K}} d\psi \wedge d\phi \right)
\ee 
Dilaton is given by $e^{2\Phi}= \frac{Q_{5}}{Q_{1}}$. Details of final T-duality are given in the appendix. Metric, after
T-duality, is given by
\begin{displaymath}
ds^{2}= \frac{Q_{5}Q_{K}}{r \Sigma}(dr^{2}+ r^{2}d\theta^{2}) + \frac{1}{4Q_{5}\tilde{R}}d\psi^{2} + \frac{2Q_{5}Q_{K}}
{\tilde{R}}(\alpha + \tilde{R}) d\phi^{2} + \frac{2Q_{K}}{\tilde{R}R_{K}}(\tilde{R}+\alpha)d\phi d\psi -\frac{r}{Q_{1}}
dt^{2} + \frac{\Sigma}{Q_{1}}
\end{displaymath} 
\be
-\frac{1}{Q_{1}}(\tilde{R}+\alpha)dtdy + \frac{4(\tilde{R}+ \alpha)Q_{K}}{R_{K}}\sqrt{\frac{Q_{5}}{Q_{1}}}d\phi(dy-dt) + 
\frac{(\tilde{R}+ \alpha)}{2\tilde{R}\sqrt{Q_{5}Q_{1}}}d\psi(dy-dt)
\ee 
Dilaton is given by $e^{2\Phi'}= \frac{1}{4Q_{1}\tilde{R}}$. Non-zero components of B-field are given by
\begin{eqnarray}
B'_{t\psi}= \frac{\tilde{R}-\alpha}{4\tilde{R}\sqrt{Q_{1}Q_{5}}} \ \ , \ \ B'_{y\psi}= \frac{\tilde{R}+\alpha}
{4\tilde{R}\sqrt{Q_{1}Q_{5}}} \\
B'_{\phi \psi}= \frac{R_{K}\alpha}{4\tilde{R}^{2}} \ \ \ , \ \ \ B'_{ty}= \frac{\Sigma + r -\tilde{R}}{2Q_{1}}
\end{eqnarray}

Since metric after T-duality is not of same form, we see that this metric naively does not look like a bound state. For a bound state, 
one would expect just a permutation of charges under duality like done above. But since we performed a T-duality along fibre direction to permute
NS$5$ and KK$6$ using Buscher rules which as shown in \cite{Tong} are insufficient to give correct answer. So the situation remains open. The question 
which we want to discuss is whether the solution constructed in \cite{bena2} has KK monopole bound to other two charges or it just acts as a background.
Since KK monopole is much heavier than other two components, D$1$ and D$5$ branes, it may look as if acting as background for other two and difference
 might not be apparent at supergravity level. But still one would think that in the S-dual system F1-NS5-KK where at least NS5 and KK both have masses going 
like $1/g^2$ (actual masses will depend on compactification radii) , it should be possible to permute these two charges. Analysis similar to \cite{Tong}
 could be done for 3-charge system to completely fix this issue. We intend to look further in this matter in a future publication.

\section{ T-duality to KK-F1}

We now convert KK-P system to KK-F1 by T-dualizing along the $w$ direction where $v=t-w$ and $u=t+w$. We are 
in IIA supergravity since we can connect this to D1-D5 by S-duality followed by T-duality along a perpendicular direction.
Writing the metric as
\begin{displaymath}
ds^{2}= -(dt^{2}-dw^{2}) + 2C_{i}dx_{i}(dt-dw) + K(dt-dw)^{2} + H[ds + A_{j}dx_{j}- A_{j}\dot{F}_{j}dt + A_{j}\dot{F}_{j}dw]
^{2}
\end{displaymath}
\be
 + H^{-1}dx_{j}dx_{j} + dz_{l}dz_{l}
\ee 
where we have written $C_{i}= (1-H^{-1})\dot{F}_{i}$ and $K= -(1-H^{-1})\dot{F}^{2}$. From now on we will leave the trivial
 torus coordinates $z_{l}$ in what follows. 
It is easier to write the T-dual metric using
\be
ds^{2}_{T}= ds'^{2} -\frac{G_{\mu w}G_{\nu w}dx^{\mu}dx^{\nu}}{G_{ww}} + \frac{(dw + B_{\mu w} dx^{\mu})^{2}}{G_{ww}}
\ee
Since there is no $B$-field in KK-P, we get
\begin{displaymath}
ds^{2}= -dt^{2} + 2C_{i}dx_{i}dt + Kdt^{2} + H[ds + A_{j}dx_{j}- A_{j}\dot{F}_{j}dt ]
^{2}+ H^{-1}dx_{j}dx_{j} + dz_{l}dz_{l}
\end{displaymath}
\be
+ \frac{dw^{2} - [(K+ H(A_{l}\dot{F}_{l})^{2})dt + H(A_{l}\dot{F}_{l})ds]^{2}- (HA_{i}A_{l}\dot{F}_{l}-C_{i})
(HA_{j}A_{l}\dot{F}_{l}-C_{j})dx^{i}dx^{j}}{[1+ K +H(A_{l}\dot{F}_{l})^{2}]}
\ee 

\newsection{Conclusion}

We summarize our results and look at possible directions for future work.
\subsection{Results}
We have found gravity solutions describing multiple KK monopoles carrying momentum by applying the solution-generating
transform of Garfinkle and Vachaspati and also by using various string dualities on the known two charge solutions. The second method only yields 
smeared solution which are logarithmically singular. One important feature of these solutions is that adding momentum to multiple  KK monopoles
leads to the separation of previously coincident KK monopoles. Hence orbifold type singularities of coincident KK monopoles are resolved and the solution is
 smooth. One can also superpose a continuum of KK monopoles carrying momentum and replace the summation by  an integral. Doing this, one
gets stationary solutions (no $t$ dependence) with isometry along $y$(compact coordinate along which the wave is travelling). 
The continuous case however, turns out to be singular. Singularity occurs at the same location where it occurs in the solution obtained by applying dualities.
In the case with $y$-isometry, we also dualized it to a KK-F1 system. 

Our reasons for studying these geometries were, in part, motivated by recent work of Bena and Kraus \cite{bena1,bena2} 
in which they constructed a smooth solution carrying D1, D5 and KK charges. This solution is supposed to represent 
one of the microstates of this system. But in these solutions, KK monopole is separated from D1 and D5 branes and acts more
like a background in which the D1-D5 bound states live. One effect of this is that the system does not appear to be 
duality symmetric i.e one can not permute the charges by performing various string dualities. We performed a specific
duality sequence (in the near horizon limit, for simplicity) to permute the charges and found that the solution is 
not symmetric. But since we used only Buscher T-duality rules, our result does not conclusively show the unboundedness of the geometry and it is possible that duality rules
going beyond supergravity will restore the symmetry. Our KK-F1 geometry is also
different from the two charge geometries one would get from Bena-Kraus geometries by setting one charge to zero. 

\subsection{Future Directions}
We have constructed time-dependent KK-P geometries which are perfectly smooth and it would be interesting to study them further. On the 
microscopic side, one can perform DBI analysis on KK-brane \cite{Bergshoeff} . First thing that needs to be checked is that brane-side gives same 
value for conserved quantities like angular momentum as the gravity side. It is expected that on the microscopic side, system would be dual to 
usual supertubes. One can also do perturbation analysis on brane-side and gravity side as done 
in \cite{yks}. Perturbation calculation for other two charge systems were also done in \cite{Giusto, Mathur} and yielded results in 
agreement with microscopic expectations.  Microscopic side of KK-branes is not very well understood, as far as we know. So calculations on gravity side should give us 
information about microscopic side and vice-versa. We will carry out some of these computations in a forthcoming publication \cite{Self}. It would 
also be interesting to explore further the connection between these solutions and black rings in KK monopole backgrounds as found in \cite{elvang1} as that
might suggest ways to add the third charge to these two-charge systems.
  
\section{ Acknowledgements}

I would like to thank my advisor Samir Mathur for suggesting the problem and for several helpful discussions at various stages of the project.
I would also like to thank Stefano Giusto and Ashish Saxena for several discussions and help with the manuscript and Borun D. Chowdhury for help in making 
the diagram.

\appendix

\newsection{T-duality formulae}
\renewcommand{\theequation}{A.\arabic{equation}}
\setcounter{equation}{0}

In this paper we perform T dualities following the notation of  
 \cite{johnson}.
 Let us summarize the relevant formulae.
 We call the T-duality direction $s$. For NS--NS fields, one has
 \bea
 G'_{ss}=\frac{1}{G_{ss}},\qquad
 e^{2\Phi'}=\frac{e^{2\Phi}}{G_{ss}},&&\ G'_{\mu s}=\frac{B_{\mu
 s}}{G_{ss}},\qquad B'_{\mu s}=\frac{G_{\mu s}}{G_{ss}}
\nonumber\\
G'_{\mu \nu}=G_{\mu \nu}-\frac{G_{\mu s}G_{\nu s}-B_{\mu s}B_{\nu  
s}}{G_{ss}},
&&\
B'_{\mu \nu}=B_{\mu \nu}-\frac{B_{\mu s}G_{\nu s}-G_{\mu s}B_{\nu  
s}}{G_{ss}},
\eea
while for the RR potentials we have:
\bea\label{RRTDual1}
{C'}^{(n)}_{\mu\dots\nu\alpha s}&=&C^{(n-1)}_{\mu\dots\nu\alpha}-
(n-1)\frac{C^{(n-1)}_{[\mu\dots\nu|s}G_{|\alpha]s}}{G_{ss}},\\
{C'}^{(n)}_{\mu\dots\nu\alpha\beta}&=&C^{(n+1)}_{\mu\dots\nu\alpha\beta  
s}
+nC^{(n-1)}_{[\mu\dots\nu\alpha}G_{\beta]s}
+n(n-1)\frac{C^{(n-1)}_{[\mu\dots\nu|s}B_{|\alpha |s}G_{|\beta  
]s}}{G_{ss}}.
\eea

\newsection{ Garfinkle-Vachaspati transform}
\renewcommand{\theequation}{B.\arabic{equation}}
\setcounter{equation}{0}

Wave-generating transform found by Garfinkle and Vachaspati belongs to the class of generalized Kerr-Schild transformations.
If one has a vector field $k_{\mu}$ which has following properties
\begin{equation}
 \  k^{\mu}k_{\mu} = 0,\  k_{\mu ; \nu}+ k_{\nu;\mu} =0,\ k_{\mu;\nu}= \frac{1}{2}(k_{\mu}A_{,\nu}-k_{\nu}A_{,\mu}) 
\end{equation}
where $A$ is some scalar function and covariant derivatives are with respect to some base metric $g_{\mu\nu}$. Then one has 
a new metric 
\be
g'_{\mu\nu}= g_{\mu\nu} + e^{A}\Phi k_{\mu}k_{\nu}
\ee
which describes a gravitational wave travelling on the original metric provided matter fields satisfy some conditions and
the function $\Phi$ satisfies
\be
\nabla^{2}\Phi =0 \ , \ k^{\mu}\partial_{\mu}\Phi =0 
\ee
Nullity of the vector field allows us to `linearize' the Einstein equations. By employing additional conditions (killing,
hypersurface-orthogonality) on the vector field, Garfinkle and Vachaspati found that Einstein equations reduce to simple
 harmonicity of a scalar function and some conditions on the matter field. Authors of \cite{myers} discuss conditions on 
matter fields in the context of low energy effective action in string theory. Consider the action 
\be
S= \int d^{D}x \sqrt{-g}\left( R - \frac{1}{2}\sum_{a} h_{a}(\phi)(\nabla \phi_{a})^{2} - \frac{1}{2}\sum_{p}f_{p}(\phi)
F^{2}_{(p+1)} \right)
\ee
Here we have included a set of scalar fields $\phi_{a}$ with arbitrary (non-derivative) couplings $h_{a}(\phi)$ and $f_{p}
(\phi)$. Degree of $p$-forms appearing depends on whether we are in type IIA or type IIB theory. Since we want the vector 
field $k$ to yield an invariance of the full solution, we impose the following conditions on the matter fields
\begin{eqnarray}
L_{k}\phi_{a}= k^{\mu}\partial_{\mu}\phi_{a}=0 \\
L_{k}F_{(p+1)}= (di_{k}+ i_{k}d)F_{(p+1)}= di_{k}F_{(p+1)}=0
\end{eqnarray}

where $L_{k}$ denotes Lie-derivative with respect to vector field $k$ and $i_{k}$ denotes interior product. In the second
 equation, we have used the identity $L_{k}= di_{k}+ i_{k}d$ and also the Bianchi identity $dF_{p+1)}=0$ for forms. 
We also require a transversality condition
\be
i_{k}F_{(p+1)}= k \wedge \theta_{(p-1)}
\ee
where $p-1$ form $\theta_{(p-1)}$ necessarily satisfies $i_{k}\theta_{(p-1)}$ since $i_{k}^{2}F_{(p+1)}=0$. This 
transversality condition ensures that the operation of raising and lowering the indices does not change the $p+1$ form 
field strength. With these conditions, the matter field equations of motion remain unchanged. Hence if the set $(g,\phi_{a},
A_{p})$ is a solution to supergravity equations then so is $(g',\phi_{a},A_{p})$.  
Note are that all this is in Einstein frame but it can be rephrased in string frame very easily. Only change is that
if Einstein and string metrics are related by
\be
g^{S}_{ab}= e^{C}g^{E}_{ab}
\ee
then Laplacian condition above becomes
\be
\partial_{\mu}(e^{\frac{(2-D)C}{2}}\sqrt{g^{S}}g^{\mu\nu}_{S}\partial_{\nu}\Phi)=0
\ee
Our null, killing vector 
is $(\frac{\partial}{\partial u})^{a}$. Since nothing depends on $u,v$ and it is a light-like direction, it is obvious that
this is null and killing. Since there is no mixing between $u,v$ and other terms, this is also hypersurface orthogonal
with $e^{A}= H^{-1}$. To see this we explicitly check hypersurface-orthogonality condition. We have $k^{u}=1$ and 
$k_{v}=g_{uv}k^{u}= g_{uv}$ as the only non-zero component. We use this in the hypersurface orthogonality condition
\be
\partial_{\nu}k_{\mu} -\Gamma_{\nu\mu}^{\lambda}k_{\lambda} = \frac{1}{2}(k_{\nu}\partial_{\mu}A-k_{\mu}\partial_{\nu}A)
\ee
Now we consider various cases. We use the fact that nothing depends on $u$ or $v$. We have following connection 
components which we will need. 
\be
\Gamma^{v}_{u \nu} =0 \ , \ \Gamma^{v}_{v \nu}= \frac{1}{2}\partial_{\nu}\ln g_{uv} \ , \ \Gamma^{v}_{i v}= \frac{1}{2}
\partial_{i}\ln g_{uv}
\ee
For $\mu =u$, we see that hypersurface orthogonality condition is trivially satisfied as all the terms vanish on both sides.
For $\mu =i$, we only have non-zero terms for $\nu =v$ and  in that case
\be
-\frac{1}{2}\partial_{i}\ln g_{uv}k_{v} = \frac{1}{2}k_{v}\partial_{\mu}A 
\ee
This gives $e^{A}= g^{uv}= (g_{uv})^{-1}$. From the other case $\mu=v$, we get the same value for $A$ and hence equations
 are consistent. With this value, we get
\be
e^{A}k_{\mu}k_{\nu}dx^{\mu}dx^{\nu} = \frac{1}{g_{uv}}g_{uv}dv g_{uv}dv = g_{uv}dv^{2}
\ee
So new metric is 
\be
ds^{2}= -(dudv + Tdv^{2}) + H^{-1}dx_{i}dx_{i}+ H(ds + V_{j}dx^{j})^{2}
\ee
We need to solve Laplace equation in the Taub-NUT geometry. Since derivatives with respect to $u$ or $v$ and along torus directions 
are zero, we have only Laplace equation 
\be
\frac{1}{\sqrt{g_{TN}}}\partial_{i}(\sqrt{g_{TN}}g^{ij}\partial_{j}T)=0
\ee
 For Taub-NUT metric, we have
\begin{eqnarray}
\sqrt{g_{TN}}= H^{-1} r^{2}\sin\theta \ , \ g^{rr}= H \ , \ g^{\theta\theta}= \frac{H}{r^{2}} \ , \ \\
g^{\phi\phi}= \frac{H}{r^{2}\sin^{2}\theta} \ , \ g^{s\phi} = -\frac{HQ_{K}\cos\theta}{r^{2}\sin^{2}\theta}=g^{\phi s}
\ , \ g^{ss}=  H^{-1}+ \frac{H Q_{K}^{2}\cot^{2}\theta}{r^{2}}
\end{eqnarray}

Using these, we write down Laplace equation for $T$ as
\dm
\partial_{r}(Hr^{2}\sin\theta\frac{1}{H}\partial_{r}T) + \partial_{\theta}(Hr^{2}\sin\theta\frac{1}{Hr^{2}}\partial_{\theta
}T) + \partial_{s}(Hr^{2}\sin\theta\frac{(H^{-2}r^{2}+ Q_{K}^{2}\cot^{2}\theta}{Hr^{2}}\partial_{s}T)
\edm
\be
+  \ \partial_{\phi}^{2}(H^{-1}r^{2}\sin\theta\frac{1}{H^{-1}r^{2}\sin^{2}\theta}T) -2\partial_{\phi}\partial_{s}(H^{-1}r^{2}
\sin\theta\frac{HQ_{K}\cos\theta}{r^{2}\sin^{2}\theta}T) =0
\ee 
Dividing by $\sqrt{g_{TN}}$, we get
\dm
\frac{1}{r^{2}}\partial_{r}(r^{2}\partial_{r}T) + \frac{1}{r^{2}}\left(\frac{1}{\sin\theta}\partial_{\theta}(\sin\theta
\partial_{\theta}T) + \frac{1}{\sin^{2}\theta}(Q_{K}^{2}\partial_{s}^{2}T + \partial_{\phi}^{2}T -2Q_{K}\cos\theta
\partial_{\phi}\partial_{s}T)\right)
\edm
\be
+ \ \frac{1}{r^{2}}(H^{-2}r^{2}-Q_{K}^{2})\partial_{s}^{2}T =0
\ee
If we assume that $\partial_{s}T=0$ then we simply get three dimensional Laplace equation whose solution is given in 
the main part of the paper. 

We will also need a theorem proved in \cite{myers} which says that the scalar curvature invariants of metrics
$g_{\mu\nu}$ and $g'_{\mu\nu}$ in Garfinkle-Vachaspati transform are exactly identical.

\end{document}